\documentclass[prl,twocolumn,,aps,showpacs]{revtex4} 

\usepackage{graphicx}
\usepackage{dcolumn}
\usepackage{bm}

\begin{document}

\title{Opinion Formation in Laggard Societies}

\author{Peter Klimek$^{1}$, Renaud Lambiotte$^{2}$, Stefan Thurner$^{1,3}$}
\email{thurner@univie.ac.at} 

\affiliation{
$^1$ Complex Systems Research Group; HNO; Medical University of Vienna; 
W\"ahringer G\"urtel 18-20; A-1090; Austria \\
$^2$ GRAPES; Universit\'e de Li\`ege; Sart-Tilman; B-4000 Li\`ege; Belgium \\
$^3$ Santa Fe Institute; 1399 Hyde Park Road; Santa Fe; NM 87501; USA\\
}


\begin{abstract}
We introduce a statistical physics model for opinion dynamics on random networks where
agents adopt the opinion held by  the majority of their direct  neighbors only
if the fraction of these neighbors exceeds a certain threshold, $p_u$.
We find a transition from total final consensus to a mixed phase where
opinions coexist amongst the agents.
The relevant parameters are the relative sizes in the initial opinion  distribution
within the population and the connectivity of the underlying network.
As the order parameter we define the asymptotic state of opinions.
In the phase diagram we find regions of total consensus and a mixed phase.
As the 'laggard parameter' $p_u$ increases the regions of consensus  shrink.
In addition we introduce rewiring of the underlying network during  
the opinion formation
process and discuss the resulting consequences in the phase diagram.
\end{abstract}

\pacs{89.75.Fb, 87.23.Ge, 05.90.+m}

\maketitle
 
\section{Introduction}

Many decisions of human beings are often strongly influenced by their social surroundings, e.g. 
the opinion of friends, colleagues or the neighborhood. Only a few types of decisions in few individuals 
emerge from  absolute norms and  firm convictions which are independent of the opinion of others. 
Much more common is the situation where some sort of social pressure leads individuals to conform to a group, 
and take decisions which minimize conflict within their nearest neighborhood.   
For example, if a large fraction of my friends votes for one party, this is likely to influence my opinion 
on whom to vote for; if I observe my peers realizing huge  profits 
by investing in some stock this might have an influence on my portfolio as well; and if the fraction of 
physicist friends (coauthors) publishing papers  on networks exceeds a certain threshold, I will have 
to reconsider and do the same; the social pressure would otherwise be just unbearable. 
Lately, the study of opinion formation within societies has become an issue of more quantitative scientific interest. 
In first attempts agents were considered as sites on a lattice, and opinion dynamics was incorporated 
by the so-called voter model (VM) \cite{BenNaim96}  (only two neighbors influence each other at one timestep), 
the majority rule (MR) \cite{Krapivsky03, Mobilia03} 
(each member of a group of odd size adopts the state of the local majority), 
or the Axelrod model \cite{Axelrod97} 
(where two neighbors influence themselves on possibly more than one topic 
with the objective to become more similar in their sets of opinions). 
Imposing regular lattice structure on social environments is convenient  
however, most observed structures of real-world networks 
belong to one of three classes:  
Erd\"os-Renyi (ER)  \cite{ER}, scale-free \cite{Albert02} or small-world networks \cite{Watts98}. 
This has been accounted for the VM \cite{Sood05, Castellano05, Castellano06} as well as 
for the MR on different topologies \cite{Lambetal07, Lambiotte07,Galam99}. For a review of further efforts 
in this directions see \cite{Stauffer03, Stauffer05} and citations therein. 
Another approach to model social interaction was developed out of the notion of catalytic sets 
\cite{Hanel05}, leading to an unanimity rule (UR) model \cite{Lambiotte06} on 
arbitrary networks in an irreversible formulation.

As a realistic model for many real world situations here we present 
a reversible generalization to the UR and MR models introducing 
an arbitrary  threshold governing updates ('laggard' parameter). 
The UR and MR are extremal cases of the model. 
In \cite{Watts02}  the idea of a threshold was introduced  in the context of investigating 
the origin of global cascades in ER networks of 'early-adopters'. In contrast to this work, 
where updates were only allowed in one direction, i.e. irreversible, the following model is 
fully reversible in the sense that two opinions compete against each other in a fully symmetric way.

\begin{figure}[t]
\begin{center}
\includegraphics[height=55mm] {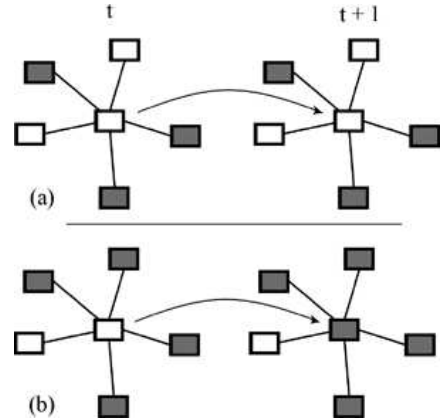}
\end{center}
\caption{Update process for two different configurations of neighbors and 
an update threshold of $p_u=0.8$. The node in the center gets updated. (a) Three out of five 
neighbors are in a different state, so the threshold is not exceeded and the node stays unchanged. 
(b) Four out of five neighbors are in a different state;  $4/5\geq0.8$  thus the node adopts the state.}
\label{update} 
\end{figure}

\section{The Model}
 
Each individual $i$ is represented as a node in a network. The state of the node represents its  
opinion on some subject.
For simplicity we restrict ourselves on binary opinions, yes/no, 0/1, Bush/Mother Theresa, etc.
Linked nodes are in contact with each other, i.e. they 'see' or know each others opinion.  
The opinion formation process of  node $i$ is a three-step process
(see Fig.\ref{update}): Suppose $i$ is initially in state '0'('1').
\begin{itemize}
\item  Check the state of all nodes connected to $i$. 
\item  If the fraction of state '1'('0')-nodes of $i$'s neighbors exceeds a threshold $p_u$, $i$ adopts opinion '1'('0').
\item Otherwise $i$ remains in state '0'('1').
\end{itemize} 

\begin{figure*}[t]
 \begin{center}
 \includegraphics[height=45mm]{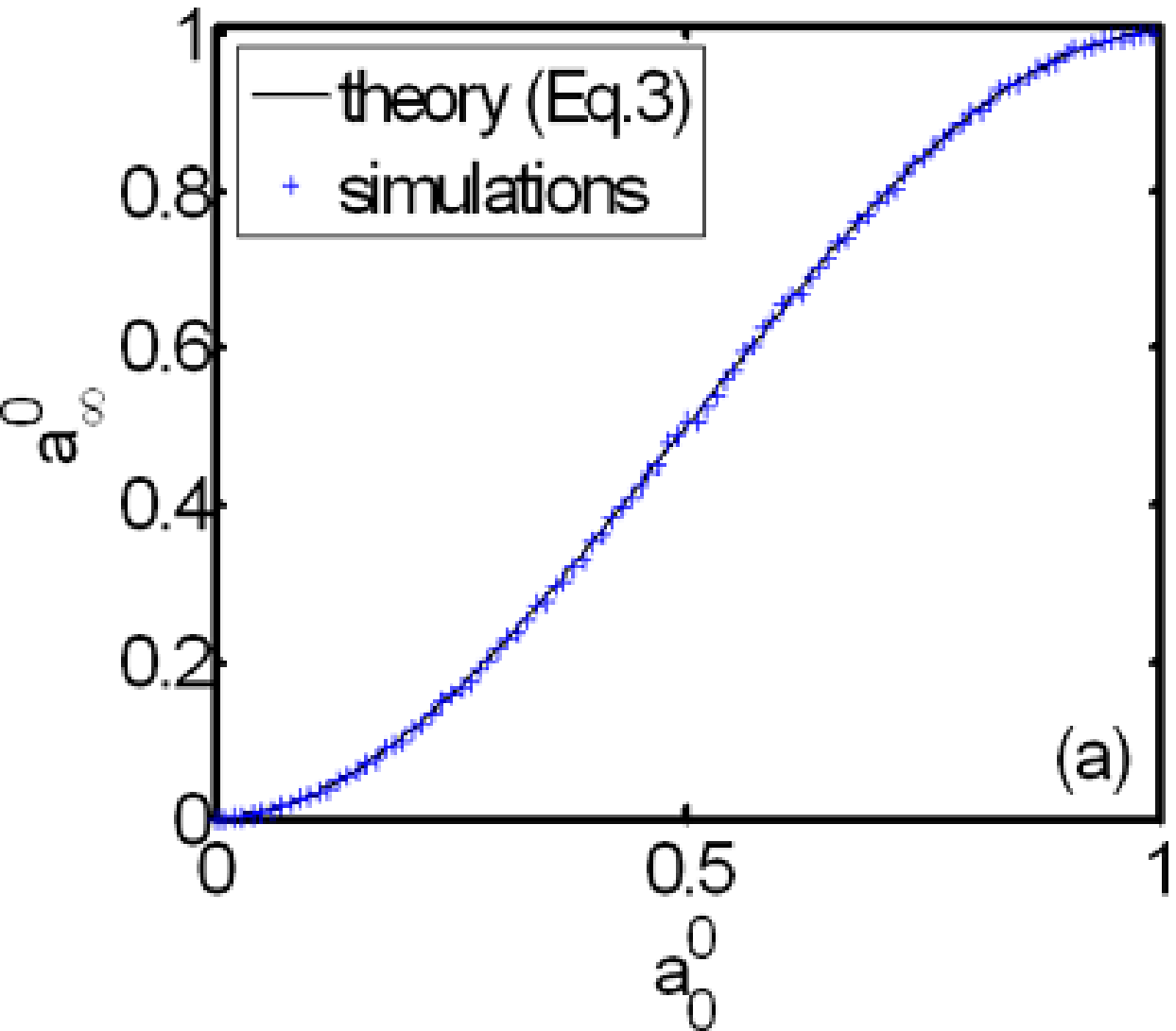}    
 \includegraphics[height=45mm]{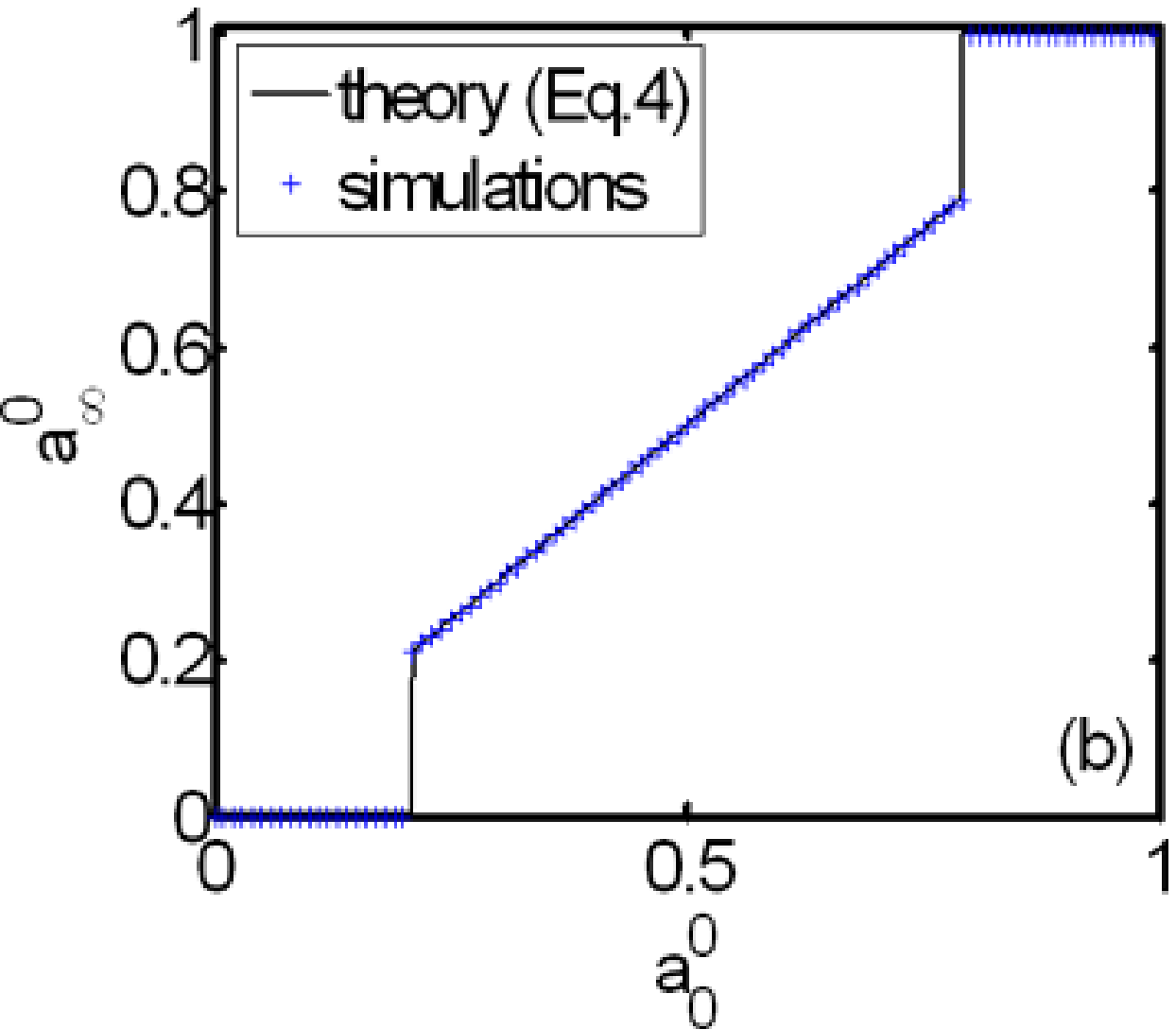} 
 \includegraphics[height=45mm]{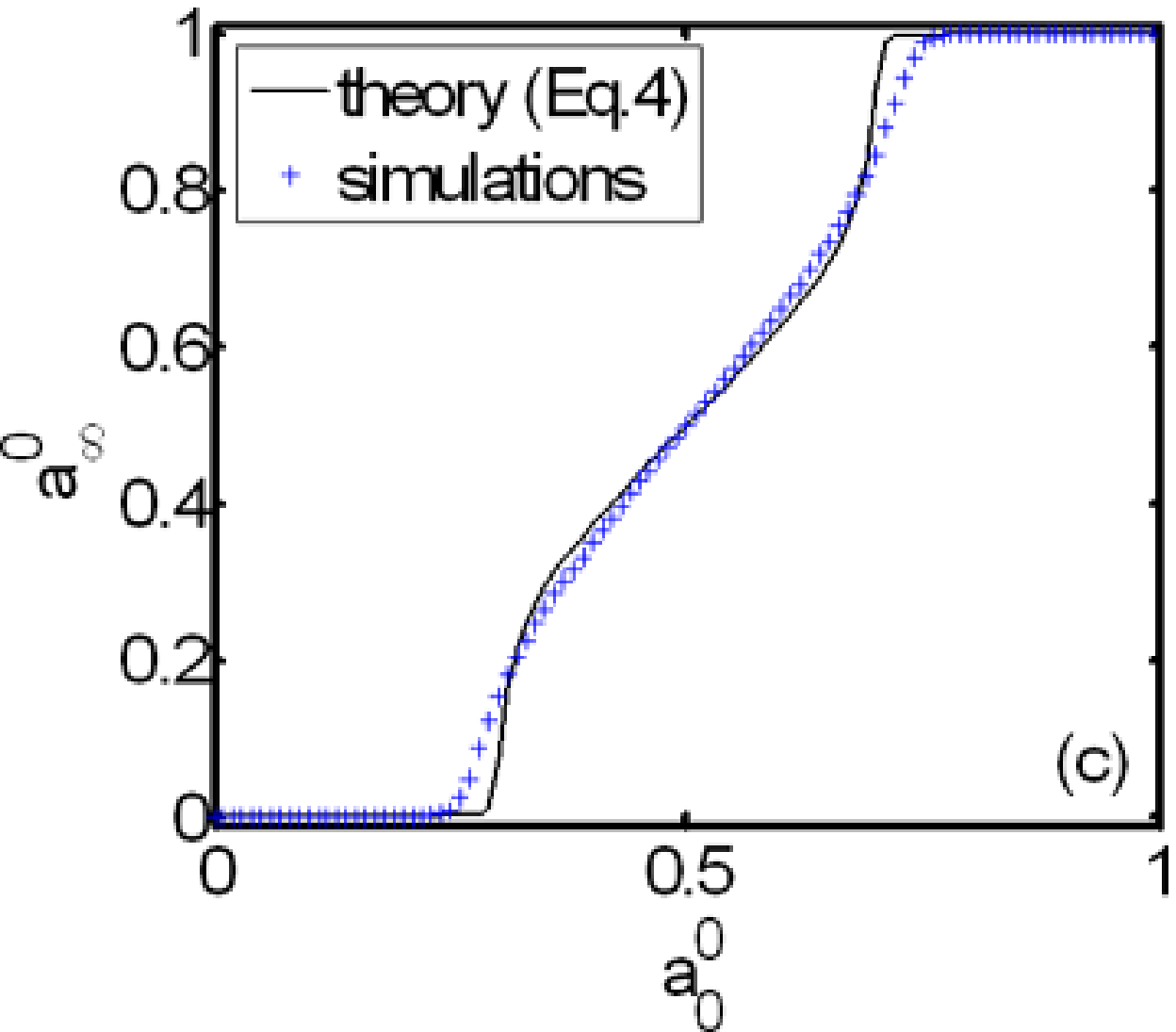}  
 \end{center}
 \caption{Asymptotic population sizes of the '0'-state fraction, $a_{\infty }^0$, as a function of its initial size, $a_0^0$,   
 for $N=10^4$, $p_u=0.8$. (a) $k=2$ for all nodes (1D circle),  (b) ER graph with $\bar{k}=9000$ and (c) 
 ER graph with $\bar{k}=10$.}
 \label{Pop}
\end{figure*}

As a substrate network we chose random graphs \cite{ER}, i.e.  
$N$ nodes are randomly linked with $L$ links (self-interactions are forbidden), 
the average connectivity being $\bar k = L/N$.   
The update threshold 
necessary for a node's change of opinion, 
$p_u$ has to be higher than $0.5$ in order 
to be meaningful in the above sense. The update is carried out asynchronously.
In a network containing $N$ nodes, at time $t$, there are $A_t^0$ nodes 
with opinion '0' and $A_t^1$ nodes with opinion '1'. The relative number 
of nodes are $a_t^{0/1}=A_t^{0/1}/N$. One time step is associated with 
applying the update procedure $N$ times, i.e. each node gets updated once per 
timestep on average. 
As time goes to infinity, the relative population of nodes with opinion 0/1 will be denoted by $a_{\infty}^{0/1}$.

\section{Analytical and Numerical Results}

To derive a \emph{master equation} for the evolution of this system we calculate 
opinion-transition probabilities via combinatorial considerations in an iterative fashion, 
motivated by \cite{Hanel05}. 
A master equation for $a_t^0$ is found explicitly, the situation for $a_t^1$ is completely analogous. 
At $t=0$, we have a fraction of $a_0^0$ nodes in state '0'. The probability that at time $t$ 
one node belonging to $a_t^0$ will flip its opinion to '1'  is  denoted by $p_t^{0 \rightarrow 1}$. 
This probability is nothing but the sum over all combinations where more than a 
fraction of $p_u$ of the neighbors are in state '1', 
weighted by the probabilities for the neighboring nodes to be either from $a_t^0$ or 
$a_t^1=\left(1-a_t^0\right)$, 
\begin{equation}
   p_t^{0 \rightarrow 1}=\sum^{\bar{k}}_{i=\lceil \bar{k} p_u\rceil} {\bar{k} \choose i} \left( 1-a_t^0 \right)^i \left( a_t^0 \right)^{\bar{k}-i} 
  \quad, 
  \label{prob}
\end{equation}
where $\lceil . \rceil$ denotes the ceiling function, i.e. the nearest integer being greater or equal. 
The same consideration leads to an expression for the opposite transition $p_t^{1 \rightarrow 0}$, where  
$1$ and $0$ are exchanged in  Eq.(\ref{prob}). 
The probability for a node to be switched from '0' to '1', 
$\Delta_0^{0 \rightarrow 1}$, is the product of the transition probability, $p_t^{0 \rightarrow 1}$, and the probability to be 
originally in the fraction $a_0^0$, i.e. $\Delta_0^{0 \rightarrow 1}=p_0^{0 \rightarrow 1} a_0^0$. 
The same reasoning gives $\Delta_0^{1 \rightarrow 0}=p_0^{1 \rightarrow 0} \left( 1-a_0^0 \right)$ and provides the master equation 
for the first time step (i.e. updating each node once on average),
\begin{equation}
  a_1^0=a_0^0+\Delta_0^{1 \rightarrow 0}-\Delta_0^{0 \rightarrow 1}  \quad. 
\label{master1}
\end{equation}
Let us now examine some special cases.

\subsection{The low connectivity limit}

For sufficiently low connectivities $\bar{k}$ there are no more possible updates 
in the network after the first iteration. For a given update threshold $p_u$, this is the case if a change in opinion requires all neighboring nodes to have the same state. To see this more clearly, consider the case 
of a network with constant $k=2$ (1D circle).  Choose a node whose state is e.g. '0'. 
There are four possible configurations of neighbors: both being in state '0',  one being 
in '0' and the other in '1' and both being in '1'. Irrespective of $p_u$, only the 
latter configuration allows an update. For all other cases at least one neighbor 
in state '0' must be updated to '1', i.e. has to have two neighboring nodes in '1'. 
But this is not possible, since at least one neighbor will always be in '0'. 
The same holds for higher values of $\bar k$, as long as every neighbor has to hold the same opinion to allow an update, i.e. we effectively use an unanimity rule.
For the special case $k=2$ the final population in state '0' is given by $a_{\infty}^{0}=a_1^0$. 
Inserting this in Eq.(\ref{master1}) yields 
 \begin{equation}
   a_{\infty}^{0}=3(a_0^0)^2-2(a_0^0)^3 \quad. 
 \label{predk2}
 \end{equation}
A comparison between the theoretical prediction of Eq.(\ref{predk2}) and the simulation of 
this system (on a regular 1D circle network with $N=10^4$) is seen 
in Fig.\ref{Pop}(a).

\subsection{Higher connectivities}

For higher connectivities there are much more configurations allowing for potential updates, 
the evolution does not stop after one single iteration. At the second time step the update 
probability is given by the product of the transition probability at $t=1$ and the probability 
to be initially in the respective state, reduced by the probability to already have undergone this 
transition during the first time step. 
We have, for example, $\Delta_1^{1 \rightarrow 0}=\left(p_1^{1 \rightarrow 0}-p_0^{1 \rightarrow 0}\right) \left( 1-a_0^0 \right)$. 
For arbitrary times $t$ this is straight forwardly seen to be 
$\Delta_t^{1 \rightarrow 0}=\left(p_t^{1 \rightarrow 0}-p_{t-1}^{1 \rightarrow 0}\right) \left( 1-a_0^0 \right)$, 
and the master equation becomes $a_{t+1}^0=a_t^0+\Delta_t^{1 \rightarrow 0}-\Delta_t^{0 \rightarrow 1}$. 
Inserting for $a_t^0$ in a recursive way yields the master equation
 \begin{equation}
   a_{t+1}^0=a_0^0+p_t^{1 \rightarrow 0}\left( 1-a_0^0 \right) - p_t^{0 \rightarrow 1} a_0^0 \quad.
 \label{master2}
 \end{equation}
Again, theoretical predictions of Eq.(\ref{master2}) agree perfectly with numerical findings, Fig.\ref{Pop}(b). 
Three regimes can be distinguished: two of them correspond to a network in full consensus. 
Between these there is a mixed phase where no consensus can be  reached.

\emph{High connectivity limit.} 
For the fully connected network the asymptotic population sizes can easily be derived: 
if $a_0^0>p_u$ or $a_0^0 < 1-p_u$ consensus is reached. For $1-p_u<a_0^0<p_u$ the 
system is frustrated and no update will take place, giving rise to a diagram like Fig.\ref{Pop}(b). 
Compared to Fig.\ref{Pop}(a) a sharp transition between the consensus phases and the mixed phase 
has appeared. We now try to understand the origin of this transition.

\begin{figure}[t]
 \begin{center}
\includegraphics[width=60mm]{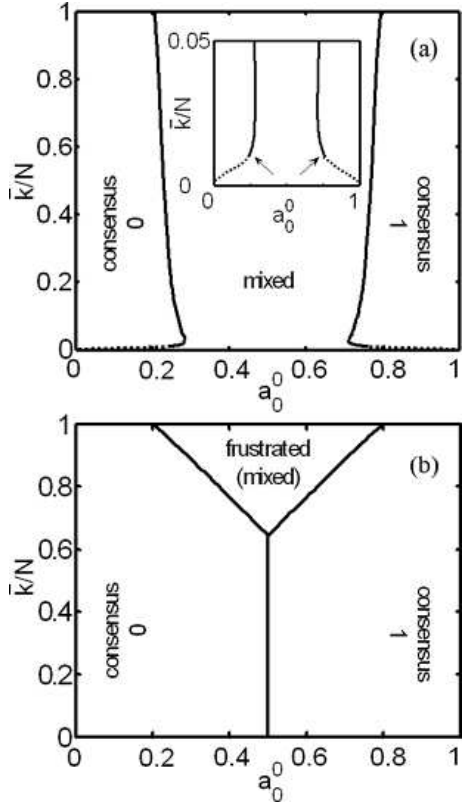} 
 \end{center}
 \caption{Phase diagram for $a_{\infty}^{0}$ as a function of initial fraction size $a_0^0$ and connectedness, $\bar{k}/N$. 
 Simulations where performed with ER graphs with  $N=10^3$ and $p_u=0.8$. Two symmetrical  regions of consensus 
 and a mixed phase in between are observed. 
 The dotted line indicates a smooth transition, the solid line a discontinuous one. 
 \emph{Inset:} Detail for small $\bar{k}$. 
 Arrows mark the change from smooth to sharp transitions, positioned at $\bar{k}/N \sim0.01$.
 (b) Phase diagram for $a_{\infty,\infty}^0$. Technically 
    adjacency matrices with $N=10^4$ were generated and checked  by Monte-Carlo simulations
    whether they allow an update at fixed $a_0^0$ and $\bar{k}$. 
}
 \label{phase}
\end{figure}

\emph{Intermediate regime.}  
The transition between the smooth solution for the final 
populations as a function of $a_0^0$ and the sharp one for higher connectivities  
becomes discontinuous when the possibility for an individual node to get updated 
\emph{in a later timestep} ceases to play a negligible role. Systems with small 
update probabilities will then be driven towards the consensus states. 
However, if the initial populations are too far from the consensus states they will not be reached.
For $p_u=0.8$ the sharp transition arises for values of $\bar{k}$ around $10$. Fig.\ref{Pop}(c) shows simulation 
data for ER graphs with  $N=10^4$ nodes and $\bar{k}=10$ with $p_u=0.8$. 
Here we already find two regimes with consensus and an almost linear regime in-between. 
The analytical curve obtained from numerical summations of Eq.(\ref{master2}) resembles the 
qualitative behavior of the simulations up to finite-size deviations. 
The dynamics of the system is shown in the phase diagram, Fig.\ref{phase}(a). 
It illustrates the size of the respective regimes and their dependence on the 
parameters $a_0^0$ and connectedness $\bar k/N$. 
The order parameter is $a_{\infty} ^0$. Along the dotted lines a smooth transition takes 
place, solid lines indicate discontinuous transitions from the consensus phase 
to the mixed phase. The change from smooth to sharp appears at 
$\bar k/N \approx0.01$. For  larger  $p_u$ the regions of consensus shrink toward the left and right margins of the figure.


So far we assumed static networks. However, this is far from being 
realistic, as social ties fluctuate.  
We thus allow links to get randomly rewired with the rewirement process 
taking place on a larger time scale than the opinion update, since otherwise 
the new connection would not lead to state changes. 
Let us assume that the number of rewired links per rewirement-timestep is fixed to $L'$, 
so that it becomes natural to define a \emph{social temperature}, $T=L'/L$.
$T$ quantifies the individual's urge to reconsider a topic with new acquaintances, or equivalently, 
the fluctuation of ties in their social surrounding.

The evolution of opinions in a network at $T\neq0$ is as follows: We fix a network and 
perform the same dynamics as for $T=0$, until the system has  converged and no further updates occur. 
Then perturb the system by a rewirement  step and randomly rewire $L'$ links among the $N$ nodes
($N$ and $L$ are kept constant over time), increase the time-unit for the rewirement steps by one 
and let the system relax into a (converged) opinion configuration. Iterate this procedure.
Note that this process can be viewed as a dynamical map of the curves shown in Figs.\ref{Pop}(a)-(c). 
With this view it becomes intuitively clear that consensus will be reached for a wider range of parameters, 
where the time to arrive there crucially depends on the value of $\bar k$. 

To incorporate the temperature effect in the master equation we introduce the second timescale 
and denote the population in state '0' as $a_{t,\bar t}^0$. Here $t$ is the time for the update process 
as before and 
$\bar t$ is the time step on the temperature time scale, i.e. counts the number of  rewirement steps.
We use $a_0^0 \equiv a_{0,0}^0$. 
$a_{\infty,0}^{0}$ can be obtained from 
$a_{\infty,0}^0=\lim_{t \rightarrow \infty} \left( a_{t,0}^0+\Delta_{t,0}^{1 \rightarrow 0}-\Delta_{t,0}^{0 \rightarrow 1}\right)$ 
for high $\bar{k}$,  and  from Eq.(\ref{master1}) for low $\bar{k}$, when we only observe updates during the first iteration. 
This evolution is nothing but a dynamical map.
The probabilities to find a configuration of neighbors allowing an update are no longer 
given only by $\Delta_{t,0}^{0 \rightarrow 1}$ and $\Delta_{t,0}^{1 \rightarrow 0}$, instead we have to count the 
ones constituted by a rewiring, which happens with probability $T$. That is why we can 
consider this kind of evolution as a dynamical map of the former process, with $a_{\infty,0}^0$ 
as the initial population for the first rewirement step evolving to $a_{\infty,1}^0$, and so on. 
The transition probabilities are now given by $T \Delta_{t,\bar t}^{1 \rightarrow 0}$ and $T \Delta_{t,\bar t}^{1 \rightarrow 0}$, 
since only new configurations can give rise to an update. We thus assume the master equation 
for a system at $T\neq0$ after the first rewiring to be
 \begin{equation}
    a_{\infty, \bar t+1}^0= \lim_{t \rightarrow \infty} \left(a_{t,\bar t}^0+ T \left(\Delta_{t,\bar t}^{1 \rightarrow 0}-\Delta_{t, \bar t}^{0 \rightarrow 1}\right) \right)
    \quad .
 \label{mastert}
 \end{equation}
Furthermore, one expects the existence of  a critical value $k_c$, below which  the intermediate regime (mixed state) 
will disappear. This will occur whenever there is no chance that a configuration of neighbors 
can be found leading to an update.
\begin{figure}[t]
 \begin{center}
 \includegraphics[height=60mm]{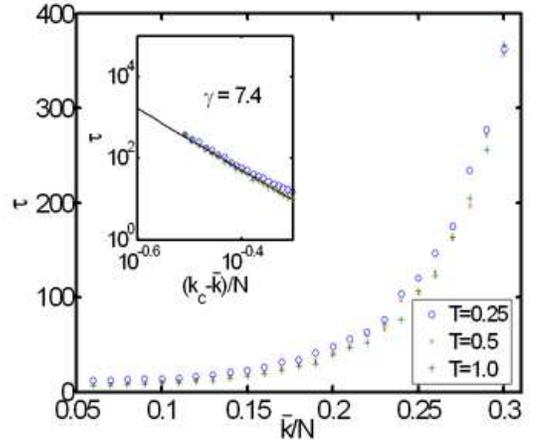}
 \end{center}
 \caption{Half-life time $\tau$  ($\bar t$ until half the population reached consensus)  
vs. relative number of neighbors for $T=0.25,0.5,1$. 
\emph{Inset:} Same in log-log scale. Scaling around the pole $k_c/N\sim0.61$
with an exponent $\gamma \approx 7.4$ is suggested. 
$10^3$ initial populations with $a_0^0=0.5$ and $N=10^2$ were averaged.
}
\label{tauvsk}
\end{figure}
The value for $k_c$ can be easily estimated: Say we have a node in state '1' and ask if an update to 
state '0' is possible under the given circumstances. For a given $\bar{k}$ this requires that there are at least 
$\lceil \bar{k} p_u \rceil$ neighbors in state '0' present in the set $A_0^0$. 
If $\bar{k}$ is above the critical value $k_c$ it occurs that even if all nodes from $A_0^0$ were  
neighbors of the node in state '1', there are still too many other neighboring nodes 
(which are then necessarily in state '1') to exceed the update threshold. 
This means that we can not have updates if $\lceil \bar{k} p_u \rceil > A_0^0$, and
we get  
 \begin{equation}
  k_c=\frac{ a_0^0 N}{p_u} \quad .
 \end{equation}
For $p_u=0.8$ and $a_0^0=0.5$, $k_c \approx 0.61 N$.
Ê
We next consider the time-to-convergence in the system. To this end we measure 
the half-life time $\tau$, of initial populations at $a_0^0=0.5$ for different connectivities 
$\bar{k}$,  see Fig.\ref{tauvsk}. 
The figure suggests that the observed scaling of $\tau$ could be of power-law type, with a pole at $k_c/N$, 
i.e. $\tau \propto \left(\frac{k_c-\bar{k}}{N}\right)^{-\gamma}$. The estimated critical exponent $\gamma \approx 7.4$ 
seems to be independent of  temperature. Note, that the estimate is taken rather far from the pole 
at $k_c$, which suggests to interpret  the actual numbers with some care.

The phase diagram for the $T\neq0$ system is shown in  
Fig.\ref{phase}(b). 
There are still three regimes, 
which are arranged in a different manner than before.  
Consensus is found for a much wider range of order parameters;  the mixed phase is found for high connectivities, i.e.  
$\bar{k}>k_c$. 
The value of $k_c$ at $a_0^0=0.5$, as found in 
Fig.\ref{phase}(b),  
is $0.63$, 
slightly above the prediction of $0.61$. 
This mismatch is  because we used networks with inhomogeneous degree distributions (Poisson). 
Whether a network allows for an update or not is solely determined by the node 
with the lowest degree $k$, which explains why we can still observe updates when 
the average degree $\bar{k}$ is near to but already above $k_c$. 
Systems in the mixed phase are frustrated. 
$k_c$ is linear in $a_0^0$ which we confirm by finding a straight line separating the frustrated drom the 
consensus phase, see 
Fig.\ref{phase}(b). 
For larger $p_u$ the regions of consensus shrink. 

\section{Conclusion}

Summarizing we presented a model bridging the gap between existing 
MR and UR models. Opinion dynamics happens 
on static random networks where agents adopt the opinion held by the majority 
of their direct neighbors only if the fraction of neighbors exceeds a pre-specified 
laggard-threshold, $p_u$. 
The larger this parameter the more stimulus the agent needs to adapt his opinion to the one of 
his direct neighborhood. 
This system shows two phases, full consensus and a 
mixed phase where opinions coexist.
We studied the corresponding phase diagram as a function of  the 
initial opinion  distribution and the connectivity of the underlying networks.
As the laggard-parameter $p_u$ increases the regions of full consensus  shrink.
We introduced rewiring of the underlying network during the opinion formation
process and discuss the resulting consequences for the phase diagram.
This social temperature introduced here differs from the usual temperature of statistical 
mechanics. It accounts for link fluctuations and not for the fluctuations of the state of the nodes. 
For $T>0$, the system can escape the frozen state $a_{\infty}^{0/1} \neq 1$, and global consensus can be obtained. 
In the case of usual temperature (opinions of nodes switch randomly) 
\cite{Lambiotte07},  a different behavior is expected. For low temperature, 
the system also can escape the frozen state, however for higher values of $T$ the system 
undergoes a transition from an ordered to an unordered phase, where $a_{\infty}=1/2$. 
Even though laggards sometimes enjoy a bad reputation as being slow and backward-oriented, 
societies of laggards are shown to have remarkable levels of versatility as long as they are not forced to 
interact too much.  

Supported by Austrian Science Fund FWF Projects P17621 and P19132 and  COST P10 action.

\end{document}